\documentclass{article}

\usepackage[english]{babel}
\usepackage{booktabs}
\usepackage{tabularx}
\usepackage{longtable}
\usepackage{siunitx}
\usepackage{enumitem}
\usepackage{xcolor}
\usepackage{amsmath}
\usepackage{authblk} 
\usepackage{newunicodechar}
\newunicodechar{−}{-}
\newcommand{\cmark}{\ding{51}}  % check mark
\newcommand{\ocirc}{\ding{109}} % open circle
\providecommand{\keywords}[1]{%
  \par\noindent\textbf{Keywords: }#1\par
}

\usepackage{graphicx} % for \resizebox (optional)
\usepackage[table]{xcolor}
\definecolor{headGrey}{gray}{0.85}

\usepackage[T1]{fontenc}
\usepackage{pifont} % dingbats
\usepackage[letterpaper,top=2cm,bottom=2cm,left=3cm,right=3cm,marginparwidth=1.75cm]{geometry}

\usepackage{comment}

\usepackage{amsmath}
\usepackage[colorlinks=true, allcolors=blue]{hyperref}
\hypersetup{
  colorlinks=true,
  linkcolor=black,
  urlcolor=blue
}

\title{Resilience by Design: A KPI for Heavy-Duty Megawatt Charging}

\author[1]{Sonia Yeh$^{*,}$}
\author[2]{Rishabh Ghotge}
\author[3]{Yujia Shi}
\author[4]{Luka de Koe}

\affil[1,3]{Chalmers University of Technology, Gothenburg, Sweden}
\affil[2,4]{Cenex, Amsterdam, North Holland NL}
\affil[*]{Contacting author: sonia.yeh@chalmers.se}
\date{January 2026}

\begin{document}
\maketitle

\begin{abstract}

We introduce a stressor-agnostic Resilience Key Performance Indicator (Resilience KPI) for megawatt charging stations (MSC) serving heavy-duty vehicles. Beyond routine performance statistics (e.g., availability, throughput), the KPI quantifies a site’s ability to anticipate, operate under degradation, and recover from disruptions using observable signals already in the framework—ride-through capability, restoration speed, service under N−1, expected unserved charging energy, and queue impacts. The headline score is normalised to 0–100 for fair cross-site and cross-vendor benchmarking, with optional stressor-specific breakouts (grid, ICT, thermal, flooding, on-site incidents) for diagnostics and robustness checks. DATEX~II provides a solid baseline for resilience KPIs centred on infrastructure inventory, status, and pricing, while additional KPIs—especially around grid capacity, on-site flexibility, heavy-vehicle geometry, environmental hardening, maintenance, and market exposure—are essential for a complete resilience picture and will require extensions or complementary data sources. The KPI is designed for monthly/quarterly reporting to support design and operational decisions and cost–benefit assessment of mitigations (e.g., backup power, spares, procedures). It offers a consistent, transparent methodology that consolidates heterogeneous logs and KPIs into a single, auditable indicator, making resilience comparable across sites, vendors, and jurisdictions.

\end{abstract}

\keywords{Resilience indicators, Electric truck charging infrastructure, Grid capacity constraints, Queuing theory, Infrastructure hardening}

\section{Purpose}

%\paragraph{Why a Resilience KPI?}
Pilot sites generate heterogeneous signals that are difficult to compare across vendors, sites, and countries. A single headline indicator enables fair benchmarking across sites, vendors, and jurisdictions, and supports longitudinal tracking over time. It turns detailed evidence into decision-grade insight for planning, procurement, and regulation—making trade-offs visible (e.g., backup power vs.\ spares vs.\ operational procedures). The Resilience KPI consolidates the Evaluation Framework’s rich data logs and KPI clusters (e.g., economics, performance, accessibility, quality of service, safety) into one normalised (0–100) indicator. Our aim is to define a single, comparable, \emph{stressor-agnostic} Resilience Key Performance Indicator (Resilience KPI) for megawatt charging stations (MCS). 

We capture resilience by combining \emph{reliability} (continuity and avoidance of interruptions), \emph{robustness} (ability to deliver service under degradation, including $N\!-\!1$ operation and thermal derating), and \emph{recovery} (detection and restoration speed, MTTR), alongside user impact (e.g., P95 start-to-charge time) and continuity of throughput (served energy ratio, expected unserved charging energy).

%\paragraph{How it integrates with the Framework}
The Resilience KPI is a \emph{compound} indicator that reuses existing measurements and ownership lines in the Evaluation Framework; no new sensors are required. It provides a single stressor-agnostic headline score for reporting and tendering, with optional stressor-by-stressor breakouts (grid, ICT, thermal, flooding, on-site incidents) for robustness checks and root-cause analysis. Results are intended for monthly/quarterly dashboards, cross-pilot comparison, and cost–benefit assessments of resilience investments. 

Optional stressor-specific breakouts (grid, ICT, thermal, flooding, on-site incidents) are retained for
diagnostics and robustness checks, while the headline indicator remains stressor-agnostic for primary
reporting.

Pilot sites produce rich but heterogeneous signals that are hard to compare across vendors, sites, and countries. The Resilience Key Performance Indicator (Resilience KPI) provides a single, stressor-agnostic headline measure (0–100) that makes resilience \emph{comparable} and \emph{actionable}. It condenses evidence already captured in the Evaluation Framework—availability and fault logs, ride-through behaviour, $N\!-\!1$ operation, restoration speed, price dynamics, and user impact (e.g., P95 start-to-charge time)—into one auditable indicator.

The objective is twofold: (i) enable fair benchmarking and longitudinal tracking; and (ii) turn detailed logs and KPIs (economics, performance, accessibility, quality of service, safety) into decision-grade insight for planning, procurement, and regulation—surfacing trade-offs such as backup power vs.\ spares vs.\ operational procedures. The Resilience KPI complements, rather than replaces, the framework’s detailed KPIs by establishing a transparent, consistent methodology that defines resilience and makes it measurable across contexts.

\section{Scope and Context}
The Resilience KPI applies to the full megawatt charging station (MCS) stack covered by the Evaluation Framework: grid interface and protection, site conversion and distribution, dispensers and connector cooling, ICT and protocol layers (e.g., ISO~15118–20/OCPI), site control/EMS, and optional BESS/backup. The primary output is a stressor-agnostic headline score (0–100) with optional stressor-specific diagnostics (grid, ICT, thermal, flooding, on-site incidents) for robustness checks and root-cause analysis.

The KPI is a compound indicator built from existing framework signals so that detailed evidence (availability/fault logs, ride-through behaviour, $N\!-\!1$ operation, recovery speed, price dynamics, user impact) is translated into a single, auditable measure. Normalisation ranges, weights, and sensitivity analyses are published to keep the score transparent and policy-ready. Results can be computed monthly or quarterly.

Computation and reporting are supported at any of the five hierarchy levels—\emph{connector}, \emph{point}, \emph{station}, \emph{pool}, and \emph{site}—with consistent roll-ups. The KPI supports single-site reporting as well as portfolio benchmarking across vendors and jurisdictions.

Where available, the DATEX~II split between static \emph{TablePublication} and dynamic \emph{StatusPublication} supports reproducible calculation and cross-site benchmarking. 
\begin{itemize}[leftmargin=1.2em]
    \item The static profile supplies site/station/\texttt{refillPoint} identifiers, geographic referencing, connector types and counts, charging modes, maximum power, payment options, \emph{ElectricEnergyMix}, and operator metadata. 
    \item The dynamic profile supplies per-\texttt{refillPoint} availability state, fault type/reason, incident timestamps (start and restoration), temporary operating hours, and current \emph{Rates}/\emph{EnergyPricingPolicy}. Together these feeds enable computation of existing KPIs K1–K14 and the composite K15, and provide all inputs required for the Resilience KPI; optional SCADA/EMS and queueing feeds can be integrated for heavy-duty operations where available.
\end{itemize}

Lightweight cyber KPIs can be derived from operator tooling—CSMS/OCPP logs, certificate/PKI dashboards, ticketing/CMMS, and asset inventories—covering link-health (LKFR, CTR, SSES), certificate health and deployment latency (CDL), patch latency, security MTTD/MTTR, and vulnerability closure rate. These integrity and recovery signals may be included in K15 with modest weights (see Section~\ref{sec:k15-construction}).

Results are suitable for single-site reports and for portfolio dashboards, cross-site comparison, and cost–benefit assessment of mitigations. Ownership follows the Evaluation Framework’s governance; thresholds, weights, and any extensions (EMS/SCADA, reservations, cyber telemetry) are documented so computation remains transparent.

\section{KPI~$\leftrightarrow$~DATEX~II Mapping}

The DATEX~II page \url{https://docs.datex2.eu/levels/mastering/energy/} defines the data model for energy supply points (e.g., EV charging, alternative fuels). Energy-resilience KPIs should be computable primarily from two publications:
\begin{itemize}[leftmargin=1.2em]
  \item \textbf{EnergyInfrastructureTablePublication} --- relatively static facts (where, what, how it is equipped).
  \item \textbf{EnergyInfrastructureStatusPublication} --- dynamic status (availability, faults, price changes, temporary hours).
\end{itemize}

The model separates static vs.\ dynamic information, enabling benchmarking (from \emph{TablePublication}) and real-time disruption scoring (from \emph{StatusPublication}). It formalizes three core classes:
\begin{itemize}[leftmargin=1.2em]
  \item \textbf{EnergyInfrastructureSite}
  \item \textbf{EnergyInfrastructureStation}
  \item \textbf{RefillPoint} (for EVs: \textbf{ElectricChargingPoint})
\end{itemize}
These inherit common \emph{Facility} properties (IDs, hours, operator/owner, rates, URLs), enabling consistent KPI roll-ups from point $\rightarrow$ station $\rightarrow$ site $\rightarrow$ region. Connector/charging definitions align with industry standards (IEC~61851, ISO~15118, IEC~62196, etc.). EU open-data profiles (RTTI/MMTIS) cover recharging point locations, availability, and ad-hoc price, facilitating interoperability.

\setlength{\extrarowheight}{2pt}
\renewcommand{\arraystretch}{1.2}

\setlength{\extrarowheight}{2pt}
\renewcommand{\arraystretch}{1.2}

\setlength{\extrarowheight}{2pt}
\renewcommand{\arraystretch}{1.2}

\begingroup
\setlength{\LTleft}{0pt}
\setlength{\LTright}{0pt}
\small % or \footnotesize / \scriptsize
\renewcommand{\arraystretch}{1.05} % (optional) tighten rows a touch
\begin{longtable}{p{0.04\linewidth} p{0.25\linewidth} p{0.33\linewidth} p{0.06\linewidth} p{0.23\linewidth}}
\caption{KPI $\leftrightarrow$ DATEX~II Mapping (Static $\rightarrow$ Static{+}Dynamic $\rightarrow$ Dynamic; overall index last).}
\label{tab:kpi-datex-mapping}\\
\toprule
\textbf{ID} & \textbf{Resilience KPI} & \textbf{DATEX~II elements} & \textbf{S/D} & \textbf{Compute / Formula} \\
\midrule
\endfirsthead
\toprule
\textbf{ID} & \textbf{Resilience KPI} & \textbf{DATEX~II elements to use} & \textbf{S/D} & \textbf{How to compute / Formula} \\
\midrule
\endhead

\multicolumn{5}{l}{\textbf{A. Static (S)}}\\
\midrule
K1 & Redundancy at site 
   & Site/Station/RefillPoint hierarchy in \emph{EnergyInfrastructureTablePublication} 
   & S 
   & $R_{\mathrm{site}}$; see Def.~\ref{kpi:Rsite}. \\

K2 & Power adequacy (high-power share) 
   & \emph{ElectricChargingPoint} $\to$ connectors, \emph{ChargingModeEnum}, max power (TablePublication) 
   & S 
   & $\mathrm{HP\_share}(P_{\mathrm{thr}})$; see Def.~\ref{kpi:HPshare}. \\

K3 & Green Supply Ratio (GSR) 
   & \emph{ElectricEnergyMix}, \emph{ElectricEnergySourceRatio} (TablePublication) 
   & S 
   & $\mathrm{GSR}$; see Def.~\ref{kpi:GSR}. \\

K4 & User access resilience (payment diversity) 
   & Facilities/payment options at site/station (TablePublication) 
   & S 
   & $\mathrm{UAR}$; see Def.~\ref{kpi:UAR}. \\

K5 & Spatial access index (coverage \& proximity) 
   & Location references (\emph{PointCoordinates}, \emph{NamedArea}) (TablePublication) 
   & S 
   & $\mathrm{SC}(r)$, $\mathrm{CSP}$, optional $\mathrm{SAI}$; see Def.~\ref{kpi:SC}. \\

\midrule
\multicolumn{5}{l}{\textbf{B. Static + Dynamic (S/D)}}\\
\midrule
K6 & Functional availability by connector (instantaneous) 
   & \emph{ElectricChargingPoint} $\to$ \emph{Connector} (\emph{ConnectorTypeEnum}, \emph{ChargingModeEnum}) + status (StatusPublication) 
   & S/D 
   & $\mathrm{FA}_{\mathrm{inst}}(c,t)$; see Def.~\ref{kpi:FAinst}. \\

K7 & Grid-outage tolerance 
   & \emph{EnergyInfrastructureStatusPublication} (temporary hours/overrides), \emph{EnergyPricingPolicy}; incident/outage tags (extension) 
   & S/D 
   & $\mathrm{GOT}$; see Def.~\ref{kpi:GOT}. \\

% <<< CHANGE: was K7b; promote to K8
K8 & Comms-outage service continuity (COSC)
    & \emph{EnergyInfrastructureStatusPublication} + CSMS comms incident/outage tags (extension)
    & S/D
    & $\mathrm{COSC}_{\mathrm{time}}, \mathrm{COSC}_{\mathrm{sessions}}$; see Def.~\ref{kpi:COSC}. \\

\midrule
\multicolumn{5}{l}{\textbf{C. Dynamic (D)}}\\
\midrule
% <<< CHANGE: shift all down by +1
K9  & Availability by connector ($A_{\text{conn}}$; time-weighted) 
    & Connector status events grouped by \emph{ConnectorTypeEnum} (StatusPublication) 
    & D 
    & $A_{\text{conn}}(c)$; see Def.~\ref{kpi:Aconn}. \\

K10 & Uptime, MTBF, and Failure Duration (MDF) 
    & \emph{FacilityStatus} with failure/restoration timestamps; optional repair-start (StatusPublication)
    & D 
    & Uptime, MTBF, MDF; see Def.~\ref{kpi:Uptime}. \\

K11 & Interruption Responsiveness (Recovery Time) 
    & \emph{FacilityStatus} transitions (interruption start/restore); threshold $P_{\min}$ (StatusPublication) 
    & D 
    & $\mathrm{IR}_{\min}$, $\mathrm{IR}_{\mathrm{full}}$; see Def.~\ref{kpi:SR}. \\

K12 & Price Instability/Volatility (PIV) 
    & \emph{Rates}, \emph{EnergyPricingPolicy} (StatusPublication); chosen window $W_{\mathrm{PS}}$ 
    & D 
    & $\mathrm{PIV}(t)$ over $W_{\mathrm{PIV}}$; see Def.~\ref{kpi:PIV}. \\

K13 & Price Surge Intensity (PSI) 
    & \emph{Rates} (StatusPublication); baseline $\mu_{\mathrm{ref}},\sigma_{\mathrm{ref}}$ 
    & D 
    & $\mathrm{PSI}(t)$ (standardised score); see Def.~\ref{kpi:PSI}. \\

K14 & Average Waiting Time and Utilisation 
    & Queue/reservation events, wait-zone/ANPR, connector handshake timestamps (extensions) 
    & D 
    & Median/P95 wait; $\rho$; see Def.~\ref{kpi:Queue}. \\

\midrule
\multicolumn{5}{l}{\textbf{D. Overall indicator (place last)}}\\
\midrule
% <<< CHANGE: SRS is now K15
K15 & Site Resilience Score (SRS) 
    & Combination of K1--K14 (plus weights) 
    & S/D 
    & $\mathrm{SRS}$; see Def.~\ref{kpi:SRS}. \\
\bottomrule
\end{longtable}

\section{Operational KPI Definitions}

\subsection{Notation and Assumptions}

\paragraph{Time domain.}
All KPIs are computed over a closed-open analysis window $[t_0, t_1)$ with $t_0 < t_1$ (timestamps in UTC unless stated). Durations such as $(t_1 - t_0)$ are measured in hours unless specified.

\paragraph{Events and statuses.}
Infrastructure components (site, station, or refill point) emit status events of the form
\[
\text{status}(x,t) \in \{\text{available},\ \text{occupied},\ \text{outOfService},\ \text{fault},\ \text{unknown},\ldots\}
\]
where $x$ can denote a site, a station, a refill point, or a connector (as applicable).
Let $\{[t_k^{\downarrow},\, t_k^{\uparrow})\}_{k=1}^{K}$ be the disjoint intervals during which the chosen component is in a \emph{fault} (or \emph{outOfService}) state within $[t_0,t_1)$; $K$ is the number of such fault intervals.
Here, $t_k^{\downarrow}$ is the start (fault onset) and $t_k^{\uparrow}$ is the end (restoration) of the $k$-th interval.

\paragraph{Indicator function.}
$\mathbf{1}\{\cdot\}$ equals $1$ if the condition is true and $0$ otherwise.

\paragraph{Connectors and groups.}
Let $\text{ConnectorTypeEnum}$ be the categorical set of connector types (e.g., CCS, CHAdeMO, Type 2, MCS).
For any connector type $c \in \text{ConnectorTypeEnum}$, let $\mathcal{C}(c)$ be the set of all installed connectors of type $c$ at the component of interest; $|\mathcal{C}(c)|$ is its cardinality.

\paragraph{Rates and windows.}
$\text{Rate}_{t:t+W}$ denotes the time series of ad-hoc (or effective session) prices observed in the rolling window of length $W$ (e.g., $W=24$\,h) starting at time $t$; $\mu(\cdot)$ and $\sigma(\cdot)$ denote mean and standard deviation.

\paragraph{Rolling window, samples, and summary operators.}
For any time $t$ and window length $W>0$, let $[t, t+W)$ be the rolling window.
Let $\{\text{Rate}(t_i)\}_{i=1}^{N_W}$ denote the $N_W$ price observations that fall inside $[t, t+W)$
(e.g., per-interval ad-hoc prices or effective session rates).
Define the sample mean and (unbiased) sample standard deviation used throughout as
\[
\mu\!\left(\text{Rate}_{t:t+W}\right) \;=\; \frac{1}{N_W}\sum_{i=1}^{N_W} \text{Rate}(t_i), 
\qquad
\sigma\!\left(\text{Rate}_{t:t+W}\right) \;=\; \sqrt{\frac{1}{N_W-1}\sum_{i=1}^{N_W} \left(\text{Rate}(t_i) - \mu\!\left(\text{Rate}_{t:t+W}\right)\right)^{2}}.
\]
If $N_W < N_{\min}$ (minimum required observations) or $\mu(\cdot)=0$, report $\text{PS}(t)$ as undefined (or use a documented fallback).
Optionally, robust alternatives (median and MAD, trimmed mean/std) may be used; if so, state the choice explicitly.

\paragraph{Composite indices.}
$\mathcal{S}$ is a set of normalized sub-KPIs (each scaled to $[0,1]$) used in a composite score; $w_i \ge 0$ are weights with $\sum_i w_i = 1$ unless otherwise stated.
$\text{FaultRate}$ can be defined as total fault time divided by $(t_1 - t_0)$, or as faults per 1000 operating hours (state clearly in reporting).

\paragraph{Queueing (heavy-duty nice-to-have).}
For M/M/$s$ approximations:
$\lambda$ is the arrival rate (veh/h), $\mu$ is the service rate per charger (veh/h), $s$ is the number of identical parallel chargers, and $\rho=\lambda/(s\mu) < 1$ is the utilization factor.
$P_0$ is the steady-state probability of zero jobs in system as defined in the formulas.

\paragraph{Units and rounding.}
Report time in hours (or minutes), energy in kWh, power in kW (or MW for MCS), prices in €/kWh (or local currency), and distances in km.
Round percentages to one decimal place and times to the nearest minute unless specified.

All sub-KPIs are normalized to [0,1] for aggregation in K15; weighting and aggregation are specified in Section~\ref{sec:k15-construction}.

\subsection{Formula}
% --- K1: Redundancy at site (Static) ---
\subsubsection{Redundancy at Site}
\label{kpi:Rsite}
Let $N_{\mathrm{rp}}$ be the number of independent refill points (e.g., ElectricChargingPoints) at the site and $N_{\mathrm{target}}>0$ a planning target (e.g., 4).
\begin{equation}
R_{\mathrm{site}} \;=\; \min\!\left(\frac{N_{\mathrm{rp}}}{N_{\mathrm{target}}},\,1\right).
\end{equation}
Compute per connector family if desired (e.g., CCS-only redundancy). If you know feeder independence, count only refill points on distinct feeders.

% --- K2: Power adequacy (Static) ---
\subsubsection{Power Adequacy (High-Power Share)}
\label{kpi:HPshare}
%For threshold $P_{\mathrm{thr}}$ (e.g., $150\,$kW), let $\mathcal{C}$ be all installed connectors and %$\mathcal{C}_{\ge P_{\mathrm{thr}}}=\{j\in\mathcal{C}: P^{\max}_j \ge P_{\mathrm{thr}}\}$.
%\begin{equation}
%\mathrm{HP\_share}(P_{\mathrm{thr}}) \;=\; \frac{|\mathcal{C}_{\ge P_{\mathrm{thr}}}|}{|\mathcal{C}|}.
%\end{equation}

% AFIR-aligned default thresholds for HDV DC recharging
\newcommand{\PthrEU}{1000\,\mathrm{kW}}   % AFIR Annex III: Level 1 ultra-fast DC starts at 150 kW
\newcommand{\PthrEUtwo}{750\,\mathrm{kW}}% AFIR Annex III: Level 2 ultra-fast DC at 350 kW
For MCS, we adopt tier (e.g., $\ge\!750$\,kW or $\ge\!1$\,MW) to reflect megawatt-class readiness in line with the industry's trajectory toward multi-MW systems.
For a fixed threshold $P_{\mathrm{thr}}$ (default $P_{\mathrm{thr}}=\PthrEU$), let $\mathcal{C}$ be the set of installed connectors and
$\mathcal{C}_{\ge P_{\mathrm{thr}}}=\{ j\in\mathcal{C} : P^{\max}_j \ge P_{\mathrm{thr}} \}$.
\begin{equation}
\mathrm{HP\_share}(P_{\mathrm{thr}}) \;=\; \frac{\lvert\mathcal{C}_{\ge P_{\mathrm{thr}}}\rvert}{\lvert\mathcal{C}\rvert}.
\end{equation}

% Optional tiered variant aligned to AFIR Annex III for diagnostics
\noindent\textit{Tiered diagnostic:}
\[
\mathrm{HP\_share}^{1000}=\frac{\lvert\{j: P^{\max}_j \ge \PthrEU\}\rvert}{\lvert\mathcal{C}\rvert}, \qquad
\mathrm{HP\_share}^{750}=\frac{\lvert\{j: P^{\max}_j \ge \PthrEUtwo\}\rvert}{\lvert\mathcal{C}\rvert}.
\]

% --- K3 ---
\subsubsection{Green Supply Ratio (GSR)}
\label{kpi:GSR}
GSR procurement firmness and curtailment risk can affect ride-through strategies, regulatory reporting, and stakeholder expectations. They can be measured both as a simple static GSR, and an optional dynamic variant when time-series data are available.

\textit{Static.} Share of delivered electricity from renewable sources at a charging point or site:
\begin{equation}
\text{GSR} \;=\; \sum_{s \in \mathcal{R}} \text{ratio}(s), \qquad \mathcal{R}=\{\text{renewable sources}\}.
\end{equation}
Source from \texttt{ElectricEnergyMix}/\texttt{ElectricEnergySourceRatio}. For smart charging \& secure comms enabling provenance and pricing transparency, see ISO~15118 guidance \cite{ECOSRAP2022}.

\textit{(Optional) dynamic KPI.} When dynamic mix shares and delivered energy are available, report an energy-weighted GSR over window $T$ (e.g.\ day or month):
\[
\mathrm{GSR}_{T} \;=\; 
\frac{\sum_{t \in T} r(t)\,E_{\mathrm{del}}(t)}{\sum_{t \in T} E_{\mathrm{del}}(t)}.
\]
\noindent \emph{Data:} time-varying renewable share $r(t)$ from \texttt{ElectricEnergySourceRatio} and delivered energy $E_{\mathrm{del}}(t)$ in \texttt{StatusPublication}. 
This optional KPI lets daily/seasonal charging patterns be linked to the renewable mix.

% --- K4: User access resilience (Static) ---
\subsubsection{User Access Resilience (Payment Diversity)}
\label{kpi:UAR}
Let $\mathcal{M}$ be the set of supported payment modalities (e.g., ad-hoc card, app, RFID/roaming, AFIR-compliant).
Let $p_m$ be the usage share or equal weighting for method $m\in\mathcal{M}$ with $\sum_{m}p_m=1$.
Define Shannon diversity normalized by its maximum:
\begin{equation}
\mathrm{UAR} \;=\; \frac{-\sum_{m\in\mathcal{M}} p_m \ln p_m}{\ln |\mathcal{M}|}.
\end{equation}

$\mathrm{UAR}\in[0,1]$, with $1$ = maximum diversity (all methods equally used) and $0$ = no diversity (all usage concentrated in a single method). Higher values indicate more payment diversity (greater access resilience). If no usage shares are available, set $p_m=1/|\mathcal{M}|$ (equal shares), yielding $\mathrm{UAR}=1$.

% --- K5: Spatial access / critical-site proximity (Static) ---
\subsubsection{Spatial Access Index (Coverage \& Proximity)}
\label{kpi:SC}
Let $i\in\mathcal{I}$ indexes a demand-weighted location (e.g., a grid cell, depot, or OD-centroid) with weight $w_i>0$; $\mathcal{S}$ is the set of sites; $d(i,\mathcal{S})=\min_{s\in\mathcal{S}} d(i,s)$ is the distance from $i$ to its nearest site. %The symbol $\mathbf{1}\{\cdot\}$ is an \emph{indicator}: it equals $1$ when the condition holds and $0$ otherwise.

Coverage at radius $r$:
\[
\mathrm{SC}(r) \;=\; \frac{\sum_{i\in\mathcal{I}} w_i\,\mathbf{1}\{\,d(i,\mathcal{S}) \le r\,\}}{\sum_{i\in\mathcal{I}} w_i}.
\]

$\mathrm{SC}(r)\in[0,1]$; it is the share of total demand within radius $r$ of at least one site. Higher values indicate better spatial access (e.g., $\mathrm{SC}(r)=0.8$ means 80\% of demand lies within $r$ of a site). %(Equivalently, $\mathrm{SC}(r)=\frac{\sum_{i:\, d(i,\mathcal{S})\le r} w_i}{\sum_{i} w_i}$.)

%Critical-site proximity (to nearest hospital, depot, etc.): let $d_{\mathrm{crit}}$ be the distance from the site to the nearest critical POI and $d_0>0$ a scale.
%\begin{equation}
%\mathrm{CSP} \;=\; \exp\!\left(-\frac{d_{\mathrm{crit}}}{d_0}\right).
%\end{equation}
%Optionally combine as $\mathrm{SAI}=\alpha\,\mathrm{SC}(r)+(1-\alpha)\,\mathrm{CSP}$ with $\alpha\in[0,1]$.

% --- K6: Functional availability by connector (instantaneous; S/D) ---
\subsubsection{Functional Availability by Connector (Instantaneous)}
\label{kpi:FAinst}
For connector type $c$, with installed set $\mathcal{C}(c)$ and status $\text{status}_j(t)\in\{\text{available},\ldots\}$,
\begin{equation}
\mathrm{FA}_{\mathrm{inst}}(c,t) \;=\; \frac{\sum_{j\in\mathcal{C}(c)} \mathbf{1}\{\text{status}_j(t)=\text{available}\}}{|\mathcal{C}(c)|}.
\end{equation}
This complements the time-weighted availability $A_{\text{conn}}(c)$ in Def.~\ref{kpi:Aconn}.

% --- K7: Grid-outage tolerance (S/D) ---
\subsubsection{Grid-Outage Tolerance}
\label{kpi:GOT}
Grid-Outage Tolerance (GOT) measures how much of a grid outage the site can keep operating through. It is a fraction in $[0,1]$: \textbf{1.0} means the site delivered at least a minimum viable level of service for the entire outage; \textbf{0.0} means no service at all during the outage; values in between mean it stayed up for part of the outage. If there is no BESS/backup (no islanding), the site has zero deliverable power when the grid is down, so by definition \textbf{GOT = 0}. With BESS/backup, if the system can island and supply $\ge X$\,kW to at least one connector, \textbf{GOT $>$ 0} (how high depends on how long that service is sustained).

%\paragraph{Definition.}
Let $\{[o_m^{\downarrow},o_m^{\uparrow})\}_{m=1}^{M}$ be grid-outage intervals affecting the site.
Choose a minimum viable power threshold $P_{\min}$ (e.g.\ one bay at $X$\,kW).
Define
\[
s(t)=
\begin{cases}
1, & \text{if available site power } P_{\mathrm{avail}}(t)\ge P_{\min},\\
0, & \text{otherwise.}
\end{cases}
\qquad
\mathrm{GOT}=\frac{\sum_{m=1}^{M}\int_{o_m^{\downarrow}}^{o_m^{\uparrow}} s(t)\,dt}{\sum_{m=1}^{M}(o_m^{\uparrow}-o_m^{\downarrow})}.
\]

%\paragraph{Discrete version (telemetry snapshots).}
With samples at times $t_k$ inside outage windows,
\[
\widehat{\mathrm{GOT}}=\frac{\sum_{m=1}^{M}\sum_{t_k\in[o_m^{\downarrow},\,o_m^{\uparrow})}\mathbf{1}\{P_{\mathrm{avail}}(t_k)\ge P_{\min}\}}
{\sum_{m=1}^{M}N_m},
\]
where $N_m$ is the number of samples in outage $m$.

% --- K8: Grid-outage tolerance (S/D) ---
\subsubsection{Comms-Outage Service Continuity (COSC)}
\label{kpi:COSC}

COSC quantifies continuity during comms outages and correct application/settlement of tariffs for offline sessions. 
Let $\{[u_n^{\downarrow},u_n^{\uparrow})\}_{n=1}^{N}$ denote intervals where the site loses connectivity to CSMS/OCPP.

Define
\[
s(t)=
\begin{cases}
1, & \text{if any connector can deliver service (ongoing session or locally authorised start at least } X \text{ kW)},\\
0, & \text{otherwise.}
\end{cases}
\]

\begin{equation}
\mathrm{COSC}_{\mathrm{time}} \;=\;
\frac{\sum_{n=1}^{N} \int_{u_n^{\downarrow}}^{u_n^{\uparrow}} s(t)\,dt}
     {\sum_{n=1}^{N} (u_n^{\uparrow}-u_n^{\downarrow})},
\qquad
\mathrm{COSC}_{\mathrm{sessions}} \;=\;
\frac{\#\,\text{offline sessions \emph{settled}}}{\#\,\text{offline sessions}}.
\end{equation}

An offline session is \emph{settled} if it is matched to a customer/tariff and billed (or explicitly zero-priced under a configured offline tariff) according to policy; sessions written off are \emph{not} settled.

% --- K9 ---
\subsubsection{Availability by
connector, $A_{\text{conn}}$}
\label{kpi:Aconn}
Time-weighted share of \emph{ElectricChargingPoint} with status $=$ available, grouped by \emph{ConnectorTypeEnum}.
\begin{equation}
A_{\text{conn}}(c) \;=\; \frac{\int_{t_0}^{t_1} \mathbf{1}\!\left\{\text{status}(c,t)=\text{available}\right\}\,dt}{t_1 - t_0},
\quad c \in \text{ConnectorTypeEnum}.
\end{equation}

Report $A_{\text{conn}}(c)$ as a proportion and as mean downtime per connector class.
For reliability primitives—Uptime, mean time between failures (MTBF), and mean duration of failure (MDF)—see Def.~\ref{kpi:Uptime}.

For interpretation and reporting norms on uptime versus user experience, see \cite{Veloz2022,ChargerHelp2024}.

% --- K10 ---
\subsubsection{Uptime, Mean Time Between Failures (MTBF), and Mean Duration of Failure (MDF)}
\label{kpi:Uptime}

Apply at level $L \in \{\text{connector},\ \text{point},\ \text{station},\ \text{pool},\ \text{site}\}$ with a stated service threshold (e.g., $\ge X$\,kW on one connector).
Let $[T_0,T_1)$ be the window, $\Delta T=T_1-T_0$, and $\{[f_k^{\downarrow},f_k^{\uparrow})\}_{k=1}^{K}$ the service-losing intervals.
Define total downtime $D=\sum_{k=1}^{K}(f_k^{\uparrow}-f_k^{\downarrow})$.

\[
\mathrm{Uptime}=1-\frac{D}{\Delta T},\qquad
\mathrm{MTBF}=\begin{cases}\dfrac{\Delta T-D}{K},&K>0\\ \text{n/a},&K=0\end{cases},\qquad
\mathrm{MDF}=\begin{cases}\dfrac{D}{K},&K>0\\ 0,&K=0\end{cases}.
\]

\noindent Uptime = share of time at/above threshold; MTBF = average \emph{uptime} between failures; MDF = average outage duration.

\textit{Optional MTTR (repair-phase only):}
if repair start $r_k^{\downarrow}$ is available,
\[
\mathrm{MTTR}=\frac{1}{K}\sum_{k=1}^{K}\big(f_k^{\uparrow}-r_k^{\downarrow}\big).
\]
State $L$, threshold, units, and whether planned maintenance is excluded.

% --- K11 ---
\subsubsection{Interruption Responsiveness (Recovery Time)}
\label{kpi:SR}

Average time to restore service after an interruption of service is aligned with EU regulatory terminology \cite{CEER2022Benchmarking}.
Let $k=1,\dots,K$ index service-interruption events with start $t_k^{\downarrow}$.
Define two recovery thresholds: (i) \emph{minimum viable service} $\ge P_{\min}$ (e.g., one bay at $X$\,kW); (ii) \emph{full service} (baseline capacity restored).
Let $t_{k,\min}^{\uparrow}$ and $t_{k,\mathrm{full}}^{\uparrow}$ be the first timestamps meeting each threshold. Then
\[
\mathrm{IR}_{\min}=\frac{1}{K}\sum_{k=1}^{K}\!\big(t_{k,\min}^{\uparrow}-t_k^{\downarrow}\big),
\qquad
\mathrm{IR}_{\mathrm{full}}=\frac{1}{K}\sum_{k=1}^{K}\!\big(t_{k,\mathrm{full}}^{\uparrow}-t_k^{\downarrow}\big).
\]

\noindent Report both where available; otherwise report the one computed and label it clearly. Exclude planned maintenance; censor still-open events at period end; provide $K$ and the share of events with both timestamps. Compute per stressor when tags are available. Interpret alongside uptime/availability and ride-through metrics \cite{Raoufi2020,Amiri2024,NRELResilienceBasics}.

% --- K12 ---
\subsubsection{Price Instability/Volatility (PIV)}
\label{kpi:PIV}
Rolling coefficient of variation of the ad-hoc price (or session rate) over a fixed window $W_{\mathrm{PS}}$.

\[
\mathrm{PIV}(t) \;=\; \frac{\sigma\!\big(\mathrm{Rate}_{[t,\,t+W_{\mathrm{PS}})}\big)}{\mu\!\big(\mathrm{Rate}_{[t,\,t+W_{\mathrm{PS}})}\big)}.
\]

\textit{Rule-of-thumb for $W_{\mathrm{PS}}$ :}
\begin{itemize}
  \item Sub-hourly / intraday tariffs (RTP, balancing/intraday exposure): $W_{\mathrm{PS}}=6\text{–}12\;h$ to capture intraday volatility without over-smoothing.
  \item Day-ahead hourly tariffs: $W_{\mathrm{PS}}=24\;h$ to span one full pricing horizon.
  \item Static or infrequently updated retail tariffs (incl.\ fuel analogues): $W_{\mathrm{PS}}=7\;d$ to cover weekday/weekend patterns; consider $28\text{–}49$\,d where multi-week price cycles are known.
\end{itemize}

\textit{Discrete form (sampling every $\Delta t$):}
for $H=W_{\mathrm{PS}}/\Delta t$ samples,
\[
\widehat{\mathrm{PIV}}(t_j)=
\frac{\operatorname{sd}\!\big(R_{j:j+H-1}\big)}{\operatorname{mean}\!\big(R_{j:j+H-1}\big)}.
\]

Electricity markets commonly analyse variability on intraday (sub-hourly) and day-ahead (24\,h) horizons; retail fuel markets often exhibit weekly or multi-week cycles. Align $W_{\mathrm{PS}}$ to the shortest full update/cycle you need to assess, and publish the chosen value with results.

% --- K13: Price surge intensity (Dynamic) ---
\subsubsection{Price Surge Intensity (PSI)}
\label{kpi:PSI}
Let $\mu_{\mathrm{ref}},\sigma_{\mathrm{ref}}$ be the mean and standard deviation of price over a reference baseline (e.g., last 7 days).
Define the standardized score:
\begin{equation}
\mathrm{PSI}(t) \;=\; \frac{\text{Rate}(t)-\mu_{\mathrm{ref}}}{\sigma_{\mathrm{ref}}}.
\end{equation}
Flag surges with $\mathrm{PSI}(t)>\tau$ for a documented threshold $\tau$ (e.g., $2$ or $3$).

% --- K14 ---
\subsubsection{Average Waiting Time (Queueing) and Utilization}
\label{kpi:Queue}
Average waiting time is how long a truck waits \emph{before} charging starts:
\[
\text{Waiting time for vehicle } n \;=\; t^{\text{plug}}_n - t^{\text{join}}_n,
\]
where $t^{\text{join}}_n$ is the first recorded “intent to charge” (e.g., virtual-queue join, entry into a marked waiting area, or a “charger unavailable” denial), and $t^{\text{plug}}_n$ is the first successful connector handshake. If there is no intent signal, treat pre-plug dwell as rest/other (not queueing). Report the median and P95 waiting time and the fraction of arrivals with a recorded join signal.

\textit{Utilisation.} Publish utilisation alongside waiting-time statistics:
\[
\rho \;=\; \frac{\text{arrival rate (veh/h)}}{\text{service rate per charger (veh/h)} \times s},
\]
with $s$ the number of parallel chargers. Interpret $\rho<1$ as capacity keeping up on average; $\rho\ge 1$ indicates queues will tend to grow at peak.

Compute from operator telemetry, on-site sensors (e.g., ANPR/geo-fence/wait-zone), and reservation feeds; see operational KPI practice in heavy-duty charging \cite{NACFE2023Report,NRELMDHD2025}.

If a model proxy is used elsewhere, document its assumptions (First In, First Out (FIFO) service, infinite calling population, unlimited waiting area) and input values separately.

% --- K15 ---
\subsubsection{Site Resilience Score (SRS)}
\label{kpi:SRS}
Weighted composite index using static capabilities with dynamic penalty:
\begin{equation}
\text{SRS} \;=\; \sum_{i \in \mathcal{S}} w_i \cdot \text{NormKPI}_i \;-\; w_{\mathrm{fault}}\cdot \text{FaultRate},
\end{equation}
where $\mathcal{S}$ may include redundancy, high-power share, multi-payment options, connector diversity, and proximity to critical POIs. Normalize each sub-KPI to $[0,1]$. Weights $w_i$ determined via expert elicitation or data-driven calibration. For resilience metric design principles, see \cite{Raoufi2020,Amiri2024,NRELResilienceBasics}.

\section{Constructing the Composite Site Resilience Score (K15)}
\label{sec:k15-construction}
%\paragraph{Rationale.}
K15 summarizes static capability and dynamic performance into a single, reproducible score to support benchmarking and decision-making. A common approach is a composite-index workflow with stakeholder-informed weighting. For example, Leon-Mateos et\,al.\ \cite{LeonMateos2021} uses Delphi-style expert input, normalization, and transparent aggregation. 

%\paragraph{Inputs.}
Select a set $\mathcal{S}$ of normalized sub-KPIs from K1--K13 (and optional HDV KPIs where available), each scaled to $[0,1]$ by a documented monotone transform (e.g., linear min–max against policy targets).

%\paragraph{Weights.}
Elicit non-negative weights $\{w_i\}_{i\in\mathcal{S}}$ with $\sum_i w_i=1$. Options to obtain weights include (i) expert elicitation/Delphi with consensus statistics (median; interquartile ranges), (ii) data-driven calibration (e.g., regression to historical disruption costs), or (iii) hybrid (expert priors updated with data). 

%\paragraph{Aggregation and penalty.}
Let $\text{NormKPI}_i\in[0,1]$ denote each normalized sub-KPI. Let $\text{FaultRate}\in[0,1]$ be the time share in fault or out-of-service states on the analysis window. Then
\[
\text{SRS} \;=\; \sum_{i\in\mathcal{S}} w_i\,\text{NormKPI}_i \;-\; w_{\mathrm{fault}}\,\text{FaultRate},
\]
with $w_{\mathrm{fault}}\ge 0$. Report SRS and a spider/radar plot of components.

%\paragraph{Transparency bundle.}
Publish (i) the exact normalization functions, (ii) the weight vector with elicitation method, and (iii) ablation/sensitivity results (e.g., $\pm$20\% weight perturbations). This mirrors best practice in composite-indicator construction—normalization, documented weighting, transparent aggregation, and sensitivity analysis—per the OECD/JRC guidance, and follows stakeholder-informed weighting as used in Port Resilience Index designs. \cite{NardoOECDHandbook,LeonMateos2021}. 

Where cyber telemetry is available, include two lightweight sub-indices to avoid swamping operational KPIs:
(i) a \emph{Cyber Link-Health sub-index} = mean of $\{1-\mathrm{LKFR},\ 1-\mathrm{CTR},\ \mathrm{SSES}\}$, and
(ii) a \emph{Cyber Recovery sub-index} = mean of the normalised $\{\mathrm{CDL},\ \mathrm{COSC}_{\mathrm{time}}\}$.
In K15, allocate modest weights (e.g., $w_{\text{link}}=0.05$, $w_{\text{recover}}=0.05$) and conduct $\pm20\%$ weight sensitivity.

Where cyber telemetry is available, the normalised \{LKFR, CTR, SSES, CDL, COSC\} may be included in $\mathcal{S}$ with weights elicited alongside availability and MTTR.

\section{Cybersecurity Resilience KPIs for EV Charging}
\label{sec:cyber-kpis}

Most of the cybersecurity KPIs proposed here are practical for a CPO to track and verify because they rely on systems already in place—CSMS/OCPP logs, ticketing, asset inventories/CMDB, and certificate dashboards. For pilot sites, the emphasis should be on a small number of design-phase “gates” that are easy to check (simple Go/No-Go decisions) and on operational indicators that boil down to a single number with a clear target. This keeps attention on avoiding preventable outages and restoring service quickly without requiring a dedicated cybersecurity team. The approach follows widely used guidance (NIST CSF~2.0; IEC~62443) while keeping process overhead low. \cite{NISTCSF2024,IEC62443}

In practical terms, the pilot review should ask for evidence that these gates are in place at design/commissioning and then track a compact “Day-1” set over the first months of operation. A sensible starter set is: Patch Latency (median days to deploy critical patches, target $\le 14$), Security MTTR (median hours to resolve security-tagged incidents, target $\le 4$), Certificate Health (share of chargers with valid, non-expiring certificates and no recent revocations, target $\ge 99\%$), and Vulnerability Closure Rate (high-severity findings closed within SLA per month, target $\ge 90\%$). Each of these yields a single, defensible number that can be pulled monthly from ordinary operator tooling and reported on the pilot dashboard.

\setlength{\extrarowheight}{2pt}
\renewcommand{\arraystretch}{1.2}
\begin{longtable}{p{0.23\linewidth} p{0.37\linewidth} p{0.27\linewidth} p{0.11\linewidth}}
\caption{Cybersecurity resilience KPIs for pilot sites (plain language, CPO-oriented).}
\label{tab:cyber-kpis}\\
\toprule
\textbf{KPI} & \textbf{How to compute (simple)} & \textbf{Why it matters} & \textbf{Refs.} \\
\midrule
\endfirsthead
\toprule
\textbf{KPI} & \textbf{How to compute (simple)} & \textbf{Why it matters} & \textbf{Refs.} \\
\midrule
\endhead

Patch Latency (days)
& Median days from security bulletin/release to deployment on chargers/CSMS; report by severity class (critical/high).
& Faster patching closes known holes and limits outages/abuse.
& \cite{NISTCSF2024} \\

MFA Coverage (\%)
& \(\frac{\text{\# privileged accounts with MFA}}{\text{\# all privileged accounts}}\times 100\).
& Stops easy account takeovers that can reconfigure chargers.
& \cite{NISTCSF2024} \\

Certificate Health (\%)
& Share of chargers with valid, non-expiring TLS/contract certificates and no recent revocation alerts (last \(W\) days).
& Broken/expired “ID cards” cause outages or insecure links.
& \cite{ISO15118,OCAOCPP201} \\

% <<< CHANGE: rename heartbeat-only KPI to a keep-alive composite
Link Keep-alive Failure Rate (LKFR) (\%)
& From CSMS logs, compute Heartbeat Failure Rate \(HFR=\frac{H_{\text{miss}}(\delta)}{H_{\text{exp}}}\) and Ping Failure Rate \(PFR=\frac{P_{\text{miss}}(\delta)}{P_{\text{exp}}}\); report both and the composite \(LKFR=\tfrac{1}{2}HFR+\tfrac{1}{2}PFR\).
& Separates app-level stalls (heartbeat) from transport/session drops (ping), catching issues earlier.
& \cite{OCAOCPP201} \\

Communication Timeout Rate (\%)
& \(\mathrm{CTR}=\frac{N_{\text{to}}}{N_{\text{tr}}}\) over \([t_0,t_1)\); segment by message family (e.g., \texttt{Authorize}, \texttt{StartTransaction}); exclude planned maintenance windows.
& Captures network faults/DoS symptoms that degrade service continuity.
& \cite{OCAOCPP201,ENISA2023} \\

% <<< CHANGE: replace TLS Handshake Success Rate with SSES + new CDL KPI
Secure Session Establishment Success (SSES) (\%)
& With (m)TLS enabled, \(\mathrm{SSES}=\frac{N_{\text{ok}}}{N_{\text{att}}}\); annotate failure codes (expiry, untrusted CA, cipher mismatch).
& Direct signal for secure link establishment reliability.
& \cite{ISO15118,OCAOCPP201} \\

Certificate Deployment Latency (days)
& Median time from certificate issuance/re-issue to acceptance by all chargers in scope (first success per charger).
& Measures recovery speed during cert rotations/rollovers; reduces avoidable comms outages.
& \cite{ISO15118,OCAOCPP201} \\

% <<< CHANGE: promote telemetry freshness and time-sync health
Telemetry Freshness SLA Compliance (\%)
& Share of status/health messages delivered within SLA (e.g., availability \(\le 60\)s; link-health \(\le 5\)min) over \([t_0,t_1)\).
& Timely data shortens detection/repair; stale data hides faults.
& \cite{NISTCSF2024} \\

Time-Sync Health (\%)
& Share of devices with absolute clock error \(\le 2\)s (NTP/PTP) during \([t_0,t_1)\).
& Good clocks keep event ordering/forensics accurate and KPIs meaningful.
& \cite{IEC62443} \\

Secure Firmware Adoption (\%)
& Share of chargers that both support \emph{and enforce} signed-firmware updates (policy = on).
& Prevents tampering and rollbacks to vulnerable versions.
& \cite{OCAOCPP201} \\

Security Incident MTTD / MTTR (h)
& Mean/median time to detect / recover for events tagged “security” in logs/tickets.
& The quicker you see and fix, the fewer customers are hit.
& \cite{Sanghvi2021} \\

Vulnerability Closure Rate (\%/month)
& \(\frac{\text{\# high-severity findings closed within SLA}}{\text{\# high-severity findings due this month}}\times 100\).
& Shows whether issues are actually being resolved, not just found.
& \cite{ENISA2023} \\

Network separation in place (Y/N + notes)
& Checklist: charging LAN separated; remote access via VPN; no direct internet from devices; logging enabled.
& Limits blast radius if anything goes wrong elsewhere on site.
& \cite{IEC62443} \\
\bottomrule
\end{longtable}

% <<< CHANGE: replace old equations with revised link-health and cert lifecycle formulas
\paragraph{OCPP/Transport-Layer Communication Integrity.}
Let $[t_0,t_1)$ be the analysis window and $\delta$ a lateness threshold.

\begin{align}
\textbf{Heartbeat Failure Rate (HFR)}\!:&& 
\mathrm{HFR} &= \frac{H_{\text{miss}}(\delta)}{H_{\text{exp}}}, 
\quad 0 \le \mathrm{HFR} \le 1. \\
\textbf{Ping Failure Rate (PFR)}\!:&&
\mathrm{PFR} &= \frac{P_{\text{miss}}(\delta)}{P_{\text{exp}}},
\quad 0 \le \mathrm{PFR} \le 1. \\
\textbf{Link Keep-alive Failure Rate (LKFR)}\!:&&
\mathrm{LKFR} &= \tfrac{1}{2}\mathrm{HFR} + \tfrac{1}{2}\mathrm{PFR}. \\
\textbf{Communication Timeout Rate (CTR)}\!:&& 
\mathrm{CTR} &= \frac{N_{\text{to}}}{N_{\text{tr}}}, 
\quad 0 \le \mathrm{CTR} \le 1. \\
\textbf{Secure Session Establishment Success (SSES)}\!:&& 
\mathrm{SSES} &= \frac{N_{\text{ok}}}{N_{\text{att}}}, 
\quad 0 \le \mathrm{SSES} \le 1.
\end{align}

\paragraph{Certificate Lifecycle.}
Let $\mathcal{D}$ be the set of devices requiring a certificate update at time of issuance/re-issue with timestamps $\{t^{\text{issue}}\}$; let $t^{\text{acc}}_d$ be the first successful secure session by device $d\in\mathcal{D}$ with the new cert chain.
\begin{equation}
\textbf{Certificate Deployment Latency (CDL)} \;=\; \mathrm{median}_{d\in\mathcal{D}}\left( t^{\text{acc}}_d - t^{\text{issue}} \right).
\end{equation}

\begin{align}
\text{\textbf{Telemetry Freshness SLA Compliance (TFS)}} \; &=\;
\frac{\sum_{m\in\mathcal{M}} \#\{\text{messages of } m \text{ with latency } \le \Delta_m\}}
     {\sum_{m\in\mathcal{M}} \#\{\text{messages of } m\}} \\
\text{\textbf{Time-Sync Health (TSH)}} \; &=\;
\frac{\#\{\text{devices with } |e_{\text{clock}}| \le 2\,\text{s}\}}
     {\#\{\text{devices}\}}
\end{align}

\noindent\textit{Data sources.} CSMS/OCPP logs and dashboards, certificate/PKI monitors, asset inventory/CMDB, vendor bulletins, and the ticketing system are sufficient. No personal data is required.

\paragraph{OCPP/Transport-Layer Communication Integrity.}
To complement the Day-1 set, track lightweight link-health indicators directly from CSMS/OCPP telemetry. Let $[t_0,t_1)$ be the analysis window and $\delta$ a lateness threshold (e.g., 2\,min).

\begin{align}
\textbf{Heartbeat Failure Rate (HFR)}\!:&& 
\mathrm{HFR} &= \frac{H_{\text{miss}}(\delta)}{H_{\text{exp}}}, 
\quad 0 \le \mathrm{HFR} \le 1.
\label{eq:hfr}
\\[0.4em]
\textbf{Communication Timeout Rate (CTR)}\!:&& 
\mathrm{CTR} &= \frac{N_{\text{to}}}{N_{\text{tr}}}, 
\quad 0 \le \mathrm{CTR} \le 1.
\label{eq:ctr}
\\[0.4em]
\textbf{TLS Handshake Success Rate (THSR)}\!:&& 
\mathrm{THSR} &= \frac{N_{\text{ok}}}{N_{\text{att}}}, 
\quad 0 \le \mathrm{THSR} \le 1.
\label{eq:thsr}
\end{align}

Here, $H_{\text{exp}}$ is the expected number of heartbeats (per configured interval) and $H_{\text{miss}}(\delta)$ counts missed or late ($>\delta$) heartbeats. $N_{\text{tr}}$ and $N_{\text{to}}$ are the numbers of OCPP transactions and those ending in timeout, respectively, grouped by message family when useful. $N_{\text{ok}}$ and $N_{\text{att}}$ are successful and attempted mutual TLS handshakes between chargers and CSMS. All three are single-number indicators sourced from ordinary CSMS/OCPP logs and certificate dashboards; they are implementation-agnostic and align with OCPP~2.0.1 security guidance. See also ISO~15118 and the OCA security whitepapers for certificate management and session establishment details~[25].

\section{Extensions / Profiles}
Most of the proposed KPIs depend on data that sit outside the current DATEX~II energy schema and therefore call for targeted extensions or auxiliary feeds. In practice, operators will need to integrate grid-side interfaces (DSO/TSO), site-level SCADA/EMS, reservation and dispatch systems, computerized maintenance management systems (CMMS), weather services, and fleet telematics (e.g., SOC, target energy, dwell constraints). To avoid ad hoc integrations, we recommend defining a dedicated \emph{Heavy-Duty Charging Profile} that extends DATEX~II with fields for queue and wait-time metrics, bay geometry and circulation attributes, MCS readiness and coexistence with CCS, on-site flexibility assets (BESS capacity and state, backup/islanding parameters), snapshots of power-quality indicators, and explicit maintenance/SLA descriptors. This direction is consistent with emerging guidance for MD/HD charging site design and operations \cite{NREL91571,CharINSite2025}.

Equally important is a clean interoperability layer. All extended entities should bind to stable, standard identifiers—connector and bay IDs, charger/asset IDs, and hierarchical site–station–refill point references—so that SCADA/EMS data, DATEX~II publications, and reservation/operations feeds can be joined without loss of fidelity. With consistent IDs and timestamps, operators can compute the proposed resilience KPIs reproducibly across systems and time, enabling benchmarking, incident forensics, and regulation-ready reporting.

Where data are available, the normalized HDV-specific KPIs in this section may be included in the composite input set $\mathcal{S}$ for the Site Resilience Score (K15), with weights and aggregation per Section~\ref{sec:k15-construction}.

\begin{comment}
\subsection{Queueing and Throughput}
To understand whether a site can maintain service levels during peaks or partial outages, track the distribution of queue length and wait time over both real time and history. Shorter and less volatile waits generally indicate higher resilience to demand spikes or charger downtime. 

\emph{Queueing time is defined as the interval from the first recorded charging intent (e.g., virtual-queue join, entry into a marked waiting zone, or a “charger unavailable” denial) until the first successful connector handshake; if no intent signal exists, pre-plug dwell is not counted as queueing.} Complement this with turnaround time from arrival to exit and time-to-first-charge; lower medians and high-percentile values reflect efficient layout and operations that preserve throughput when assets are constrained. Reservation reliability (the share of bookings honored, lateness tolerance, and no-show rate) reduces unplanned stacking and spillback at peaks, while a simple “stress throughput” measure—vehicles per hour at a specified \(\Delta\)SOC during peak—acts as a practical capacity proxy when a subset of chargers is down. These metrics are straightforward to compute from operator telemetry, site sensors, and reservation feeds. \cite{NACFE2023Report,NRELMDHD2025}
\end{comment}

\subsection{Grid Capacity, Power Quality, and On-Site Flexibility}
Grid-facing attributes determine how well a site rides through disturbances and how quickly it recovers. Connection capacity and real-time headroom (relative to transformer ratings and contracted limits) improve tolerance to overlapping peaks and enable staging of fast sessions. Power-quality indicators—voltage sag/swell, total harmonic distortion, and flicker—matter because poor quality increases protective trips and faults that undermine resilience. On-site flexibility, especially BESS energy/power and availability, provides limited operations during grid events and can shorten recovery time; backup and islanding capability (black-start readiness and transfer time) further supports continuity during outages. Finally, predictable participation in demand response (event counts, delivered curtailment, response latency) and any V2G/V2X support for on-site loads round out the operational picture. Most of these can be sourced from DSO/TSO interfaces, SCADA/EMS, and charger or EMS logs. \cite{NRELMDHD2025}

\subsection{Data Quality and Telemetry Fitness}
Reliable situational awareness depends on telemetry latency and availability, data completeness, and schema compliance. High-quality, timely data reduce blind spots and accelerate both fault detection and recovery, making the rest of the KPIs actionable. In practice, this view comes from CPO APIs, SCADA exports, and observability dashboards. \cite{NRELMDHD2025}

\section{Summary}
\label{sec:summary}

This section ties together the full KPI set with two complementary views:
(i) a high-level grouping by operational category (Fig.~\ref{fig:terminology}), and
(ii) how each KPI maps onto the three performance dimensions—\emph{reliability}, \emph{robustness}, and \emph{resilience}—in Table~\ref{tab:kpi-3R-matrix-ordered}.
Use the category view for ownership and workflow (who acts on what); use the 3R view to interpret behaviour under stress and recovery.

\subsection{Ownership \& Governance.} Align each category in Fig.~\ref{fig:terminology} to an accountable team (operations, grid/EMS, product/payments, finance, security). Use K15 as the site-level roll-up, with sensitivity to weights (Section~\ref{sec:k15-construction}). Escalate breaches of target thresholds via clear RACI (who fixes, who funds, who signs off).

For cyber KPIs, \textbf{Responsible}: CPO SecOps/Operations (monitor \& remediate breaches, patches, certificates); \textbf{Accountable}: CISO/CTO (targets, resources, exceptions); \textbf{Consulted}: vendor/CSMS, site IT/OT, DSO (as relevant); \textbf{Informed}: site owner, fleet customers, regulator (if required).

\begin{figure}[h!]
\centering
\includegraphics[width=0.8\linewidth]{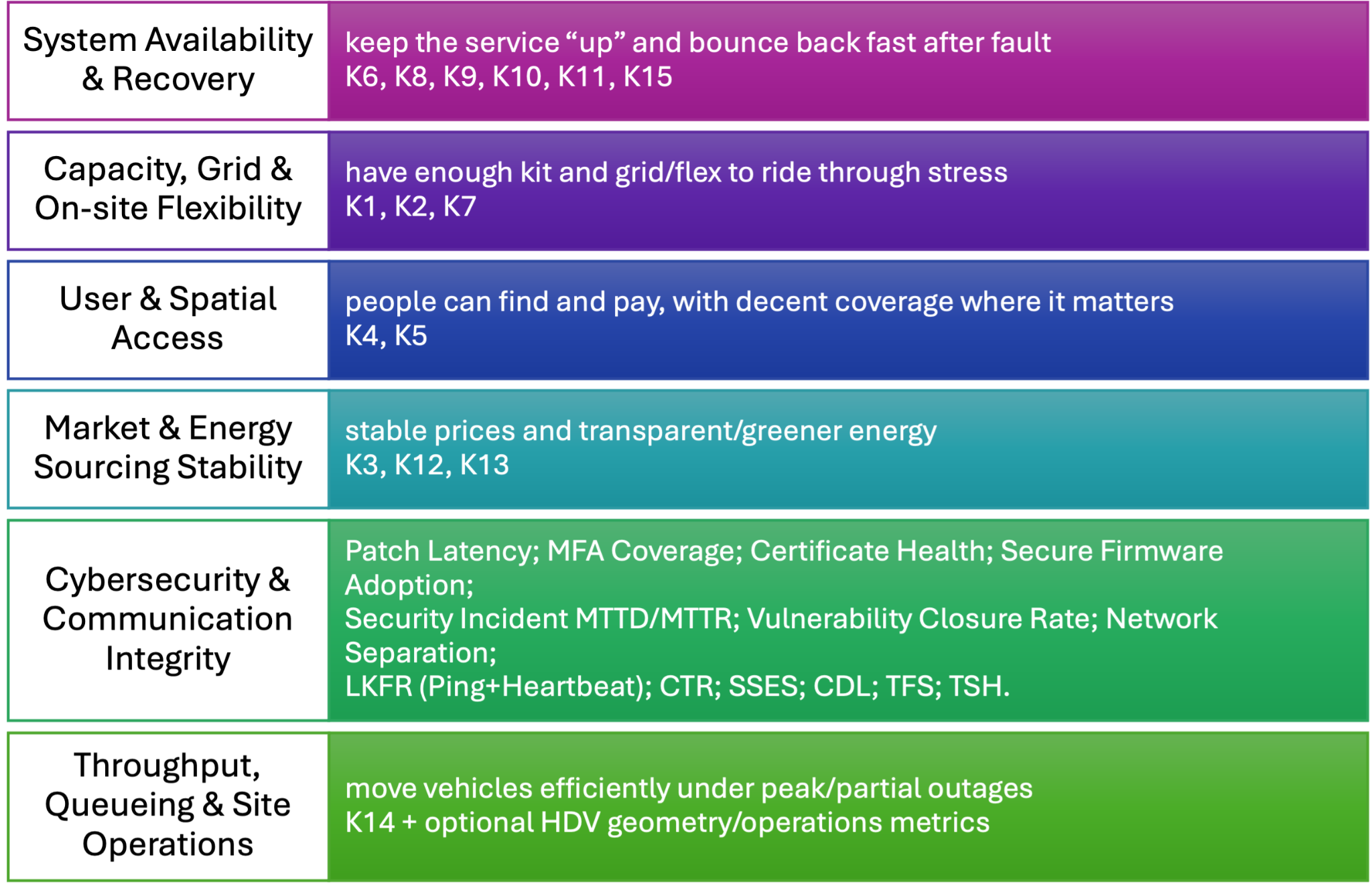}
\caption{Resilience KPI portfolio grouped into six operational categories.}
\label{fig:terminology}
\end{figure}

\subsection{Reliability, Robustness, and Recovery}
\label{sec:rrr-context}

Most KPIs in this memo are not ``resilience KPIs'' in a strict sense; they span three complementary concepts that matter for dependable charging service:
\begin{itemize}
  \item \textbf{Reliability} — ability to perform the intended function in steady state (ideally near 100\%). Typical indicators are time-in-service and fault statistics (e.g., $A_{\text{conn}}$, uptime, MTBF/MTTR).
  \item \textbf{Robustness} — ability to maintain acceptable performance under bounded disturbance without structural change (e.g., peaks, moderate link degradation). This is the ``90--100\% under stress'' band and includes capacity/redundancy, payment continuity, and live comms health (LKFR, CTR, SSES). 
  \item \textbf{Recovery} — ability to absorb shocks, adapt, and \emph{recover} functionality and quality after a disruption (speed, extent, and quality of recovery), including incident detection/response (security MTTD/MTTR) and sustained closure of issues (vulnerability closure rate), as well as composite recovery scores (K15).
\end{itemize}

These concepts overlap in practice: the same KPI can inform more than one dimension (e.g., availability contributes to reliability and is a component in resilience scoring). To keep the dashboard interpretable, we place each KPI in the dimension where its \emph{primary intent} is strongest.%

\begin{table}[h!]
\centering
\caption{Applicability of Resilience KPIs across Reliability (steady), Robustness (under disturbance), and Recovery (after disturbance). Primary intent = \cmark; secondary contribution = \ocirc.}
\label{tab:kpi-3R-matrix-ordered}
\renewcommand{\arraystretch}{1.2}
\begin{tabularx}{\linewidth}{l c c c}
\rowcolor{headGrey}
\textbf{KPI} & \textbf{Reliability} & \textbf{Robustness} & \textbf{Recovery} \\
\toprule
\textbf{K1} Redundancy at site                    & \ocirc    & \cmark    &            \\
\textbf{K2} Power adequacy                         & \ocirc    & \cmark    &            \\
\textbf{K3} Green Supply Ratio (GSR)               &           & \cmark    & \ocirc     \\
\textbf{K4} User access resilience                 & \ocirc    & \cmark    &            \\
\textbf{K5} Spatial access index                   &           & \cmark    & \ocirc     \\
\textbf{K6} Functional availability (instant)      & \cmark    & \ocirc    & \ocirc     \\
\textbf{K7} Grid-outage tolerance                  & \ocirc    & \cmark    & \ocirc     \\
% <<< CHANGE: COSC is now K8
\textbf{K8} Comms-outage service continuity        & \ocirc    & \ocirc    & \cmark     \\
\textbf{K9} Availability by connector (time-wt)     & \cmark    & \ocirc    & \ocirc     \\
\textbf{K10} Uptime / MTBF / MDF                     & \cmark    & \ocirc    & \ocirc     \\
\textbf{K11} Interruption Responsiveness (recovery time) &         & \ocirc    & \cmark     \\
\textbf{K12} Price stability (PS)                    &           & \cmark    &            \\
\textbf{K13} Price surge intensity (PSI)             &           & \ocirc    & \cmark     \\
\textbf{K14} Average waiting time \& utilisation     & \ocirc    & \cmark    &            \\
\textbf{K15} Site Resilience Score (SRS)           &           &           & \cmark     \\
\midrule
\multicolumn{4}{l}{\textbf{Cybersecurity \& communication integrity}} \\
Patch Latency                                      & \cmark    &           & \ocirc     \\
MFA Coverage                                       & \cmark    &           &            \\
Certificate Health                                 & \cmark    & \ocirc    &            \\
Secure Firmware Adoption                           & \cmark    &           & \ocirc     \\
Security Incident MTTD / MTTR                      & \ocirc    &           & \cmark     \\
Vulnerability Closure Rate                         & \ocirc    &           & \cmark     \\
Network separation in place (Y/N)                  &           & \cmark    & \ocirc     \\
Link Keep-alive Failure Rate (Ping+Heartbeat)      & \ocirc    & \cmark    &            \\
Communication Timeout Rate                         & \ocirc    & \cmark    &            \\
Secure Session Establishment Success               & \ocirc    & \cmark    &            \\
Certificate Deployment Latency                     & \ocirc    &           & \cmark     \\
Telemetry Freshness SLA Compliance                 & \cmark    & \ocirc    &            \\
Time-Sync Health                                   & \cmark    &           &            \\
\midrule
\multicolumn{4}{l}{\textbf{Nice-to-have (HDV/operations, environment, data)}} \\
Reservation reliability (\% honoured bookings)     &           & \cmark    & \ocirc     \\
Stress throughput at peak (veh/h @ $\Delta$SOC)    & \ocirc    & \cmark    & \ocirc     \\
Turnaround time $P_{95}$ (arrival$\rightarrow$exit)&           & \cmark    & \ocirc     \\
Cold-weather operability uptime (\%)               & \cmark    & \ocirc    &            \\
Power-quality incident rate (sag/swell/THD)        & \ocirc    & \cmark    &            \\
BESS-supported service during outage (h)           &           & \ocirc    & \cmark     \\
Bay geometry adequacy score (pull-through, reach)  &           & \cmark    &            \\
Data quality \& telemetry fitness (freshness/completeness) & \cmark & \ocirc & \\
\bottomrule
\end{tabularx}
\end{table}

\subsection{Interoperability}
\label{sec:interoperability}

Interoperability here means \emph{the KPIs can be computed the same way across sites, vendors, and countries} using standardised data feeds with stable identifiers and versioned schemas. This is essential for benchmarking, audits, and regulation-ready reporting.

\paragraph{What “good” looks like.}
\begin{itemize}[leftmargin=1.2em]
  \item \textbf{Standard feeds first.} Prefer DATEX~II \emph{Table/Status} for inventory/status; OCPP~1.6/2.0.1 telemetry for charger events; ISO~15118/PKI dashboards for certificates; EMS/SCADA for grid and on-site flexibility (e.g., IEC~61850). \cite{OCAOCPP201,ISO15118}
  \item \textbf{Stable identifiers.} Use persistent site, station, refill point, connector, and asset IDs across all feeds so joins are deterministic (site$\rightarrow$station$\rightarrow$refill point$\rightarrow$connector).
  \item \textbf{Schema discipline.} Version every payload; document field semantics (units, null rules, enumerations) and deprecations. Ship a JSON Schema (or XSD) for each feed.
  \item \textbf{Latency \& completeness SLAs.} Publish status latency targets (e.g., $\leq$60\,s for availability; daily for certificates); define minimum completeness for K6/K8/K9 (e.g., $>98\%$ of intervals present).
  \item \textbf{Timebase \& clocks.} Use UTC timestamps with explicit resolution; require NTP/PTP time sync for CSMS/EMS to keep event ordering consistent.
  \item \textbf{Security \& privacy.} Enforce TLS/mTLS for telemetry (per OCPP/ISO~15118); avoid personal data—KPIs need none. Limit dashboards to aggregated results.
\end{itemize}

\paragraph{Computation profile (tiers).}
Not all KPIs need all feeds. Declare what you can compute with which standards:
\begin{itemize}[leftmargin=1.2em]
  \item \textbf{Tier A (Standards-only).} DATEX~II Table/Status $\Rightarrow$ K1, K2, K4, K5, K6, K9, K10, K12, K13 (if rates are exposed), and K15 (with weights disclosed).
  \item \textbf{Tier B (Add CSMS/OCPP).} + OCPP events/logs $\Rightarrow$ K8 (comms-outage continuity), K11 (event recovery), and cyber link-health \emph{LKFR} (Ping+Heartbeat), \emph{CTR}, \emph{SSES}.
  \item \textbf{Tier C (Add EMS/SCADA/PKI).} + EMS/SCADA/PKI $\Rightarrow$ K7 (grid-outage tolerance), Certificate Health, Patch Latency, MFA Coverage, Secure Firmware Adoption, Security MTTD/MTTR, Vulnerability Closure Rate, Network Separation, plus \emph{CDL}, \emph{TFS}, \emph{TSH}.
  \item \textbf{Tier D (HDV optional).} + Reservations/yard sensors/fleet telematics $\Rightarrow$ K14 (queueing \& throughput).
\end{itemize}

\paragraph{Interoperability Readiness Index (IRI).}
Report a single number and a per-category breakdown:
\[
\mathrm{IRI} \;=\; \frac{\text{\# KPIs computable with declared standardised feeds}}{\text{total \# KPIs in scope}}\times 100\%.
\]
Also publish a vector by category (Availability, Grid/Flex, User/Spatial, Market/Sourcing, Cyber, HDV Ops) so gaps are visible.

\section{Acknowledgement}

This project has received funding from the European Union’s Horizon Europe innovation programme under the Grant Agreement No.101192466. Copyright © MACBETH Consortium, 2026. 

\renewcommand\refname{References}
\Urlmuskip=0mu plus 1mu\relax

\bibliographystyle{IEEEtran}
\bibliography{Ref}
\end{document}